# Oxygen permeability and stability in the entropy-stabilized Co-based Perovskite oxygen permeable membranes

Zaichen Xiang[1], Rui Chen,[1] Shuangyue Wang[1], Jingjun Qin,[1] Wanyi Zhang[1], Yucheng Li[1], Lingyong Zeng,[1,2] Huixia Luo[1,*]

[1]*School of Materials Science and Engineering, State Key Laboratory of Optoelectronic Materials and Technologies, Key Lab of Polymer Composite & Functional Materials, Guangdong Provincial Key Laboratory of Magnetoelectric Physics and Devices, School of Physics, Sun Yat-sen University, Guangzhou 510275, China*
[2]*Device Physics of Complex Materials, Zernike Institute for Advanced Materials, University of Groningen, Nijenborgh 4, 9747 AG Groningen, The Netherlands*

**\*Correspondence email: luohx7@mail.sysu.edu.cn (H. Luo)**

**ABSTRACT**

Oxygen transport membranes (OTMs), enabling catalytic reaction and gas separation, support crucial chemical engineering processes and decarbonization technologies, but their applications are hindered by limited oxygen permeation fluxes and inadequate long-term stability during operation. Here, a series of high-entropy perovskite oxygen transport membranes (OTMs) based on $La_{0.5}Sr_{0.5}CoO_3$ (LSC) were designed and synthesized by the simple sol-gel method. The impact of varying doping ratios on the structure, surface morphology, oxygen permeability, and stability of these high-entropy OTMs was thoroughly examined. At 950 °C, the optimal composition, $La_{0.25}Sr_{0.25}Gd_{0.2}Nd_{0.2}Pr_{0.1}CoO_3$, achieved oxygen permeation fluxes of 1.62 mL·min$^{-1}$·cm$^{-2}$ under air/He gradient and 1.46 mL·min$^{-1}$·cm$^{-2}$ under air/$CO_2$, respectively. Remarkably, all high-entropy OTMs demonstrated stable operation for over 100 h in a pure $CO_2$ environment without a significant decline in performance. This finding paves a new way to enhance the structural and oxygen permeation stability of OTMs, and further promotes the application of OTMs in oxy-fuel combustion technologies aimed at improving $CO_2$ capture and storage efficiency.

**Keywords:** High-entropy perovskite oxides; oxygen-permeable membranes; gas separation, $CO_2$ capture

# 1. Introduction

The growing frequency of weather disasters has compelled the implementation of measures to reduce $CO_2$ emissions, thereby mitigating the extensive impact of climate change on the global environment. One promising approach to achieve this is through oxy-fuel combustion, which produces high-purity $CO_2$ and enhances the economic viability of carbon capture and storage (CCS) technologies [1,2]. Among these technologies, mixed ionic-electronic conducting (MIEC) ceramic oxygen transport membranes (OTMs) have garnered significant attention from both academia and industry. These membranes are recognized for their environmental friendliness and cost-effectiveness in gas separation during oxy-fuel combustion processes [3–11].

In oxy-fuel combustion, OTMs are inevitably exposed to high-temperature environments containing $CO_2$ or other reducing gases such as $H_2$ and CO. This necessitates that OTMs possess not only excellent oxygen permeability but also tolerance to $CO_2$ or reducing environments [12–16]. Current MIEC OTM materials are primarily categorized into single-phase and dual-phase membranes. Single-phase membranes encompass perovskite, fluorite, and $K_2NiF_4$-type single-phase oxides [4]. The general chemical formula for a single-phase perovskite oxide is $ABO_3$, where A represents a rare earth or lanthanide element, and B denotes a transition metal. Single-phase perovskite materials, typically featuring cobalt in the B position, exhibit high oxygen permeability. However, their limited phase stability and mechanical strength impede their broad industrial application [7,17]. Materials incorporating alkaline earth metals like Ba and Sr are prone to reaction with atmospheric $CO_2$, forming carbonate impurities that degrade performance by obstructing oxygen-ion transport and compromising the structural integrity[18,19]. For example, $Ba_{0.5}Sr_{0.5}Co_{0.8}Fe_{0.2}O_{3-\delta}$ (BSCF)[20] experiences a decline in both structural integrity and oxygen ionic conductivity at elevated temperatures under $CO_2$ exposure, owing to carbonate formation, which leads to poor stability. Furthermore, even Ba-free compositions such as $Ln_{0.4}Sr_{0.6}Co_{0.2}Fe_{0.2}O_{3-\delta}$ (LSCF)[21] are hindered in widespread application by the high cost and scarcity of cobalt. In contrast, cobalt-free perovskite materials, such as $Gd_xBa_{1-x}FeO_3$[22] and $BaFe_{0.9}Zr_{0.05}Al_{0.05}O_{3-\delta}$[23]), offer excellent stability. However, their oxygen permeability generally falls short of industrial requirements when compared to their cobalt-containing counterparts. Furthermore, the A-site in single-phase perovskite OTMs often contains alkaline earth elements like Ba and Sr, which contribute to high oxygen permeability due to their large unit cell volume and high oxygen vacancy concentration. Nevertheless, these elements are susceptible to reacting with $CO_2$ to form carbonates, which can degrade the material structure and impede oxygen ion transport [24]. To address the limitations of single-phase perovskite membranes, there has been considerable interest and research into dual-phase oxide materials [25]. In contrast to single-phase membranes, dual-phase membranes comprise a composite of two distinct solid phases: generally, the fluorite phase acts as an oxygen-ion transporter, while the perovskite phase functions as a mixed conductor for both oxygen ions and electrons. When these two phases are homogeneously mixed, they form a dense oxygen permeation network, which facilitates the permeation and transport of oxygen ions and electrons. These phases are chemically compatible and coexist without forming a solid solution, as evidenced by distinct XRD diffraction patterns and contrasting regions in SEM-BSE imaging [26,27]. This architecture creates continuous pathways for both oxygen ions and electrons, which is essential for high oxygen permeation fluxes [28], such as $Ce_{0.9}Pr_{0.1}O_{2-\delta}$-$Pr_{0.1}Sr_{0.9}Mg_{0.1}Ti_{0.9}O_{3-\delta}$ (CPO-PSM-Ti)[29], $Ce_{0.8}Sm_{0.2}O_{2-\delta}$-$La_{0.8}Ca_{0.2}Al_{0.3}Fe_{0.7}O_{3-\delta}$ (CSO-LCAFO)[30], and $Ce_{0.8}Sm_{0.2}O_{2-\delta}$-$La_{0.8}Sr_{0.2}MnO_{3-\delta}$ (CSO-LSM) [31]. Despite improvements in $CO_2$ stability, most of these materials still do not meet the oxygen permeability requirements for practical applications [27]. To enhance oxygen permeability and maintain high stability, various improvement strategies have been proposed, including coating the membrane surface with an active catalyst [32], using a porous support layer to reduce membrane thickness [33], and improving the ionic and electronic conductivity of the fluorite phase [34–36]. Although numerous dual-phase membranes have been suggested, only a few achieve the necessary oxygen permeability to meet industrial standards (1 mL·min$^{-1}$·cm$^{-2}$). Therefore, the development of new OTMs that combine high oxygen permeability with high stability remains highly needed but still challenging.

High-entropy ceramics (HECs) offer new avenues for researchers seeking novel OTMs that combine high stability with high oxygen permeability. The fundamental characteristics of HECs include four key aspects: (1) a high entropy effect rooted in thermodynamics, (2) a structural effect involving lattice distortion, (3) a kinetic effect of hysteresis diffusion, and (4) a synergistic effect of components. This concept was initially introduced by Maria, Curtarolo, et al. in 2015 [37], and since then, HEC materials with diverse structures, such as fluorides [38], perovskites [39], and spinels [40], have been progressively developed. These materials exhibit exceptional properties and have found applications as catalysts. Drawing inspiration from research on high-entropy oxides, we propose that the precipitation of Sr and phase transformation of $La_{0.5}Sr_{0.5}CoO_3$ in high-temperature and $CO_2$ environments could be mitigated by the stability conferred by high-entropy oxides, thus improving the stability of the membrane [41,42]. Recently, Zhu et al. developed a series of $Ba_{0.5}Sr_{0.5}Co_{0.8}Fe_{0.2}O_{3-\delta}$ -based high-entropy perovskites (HEBSCF) that substantially improve the intermediate-temperature stability of BSCF[43]. Inspired by $La_{0.2}Sm_{0.2}Gd_{0.2}Nd_{0.2}Y_{0.2}CoO_3$[44], we aimed to explore the performance of the A-site high entropy cobalt-based perovskite oxygen permeable membrane. Furthermore, studies on lanthanide-doped perovskite oxygen-permeable membranes have revealed that partial substitution of $Sr^{2+}$ with lanthanides enhances ionic conductivity and stability in perovskites [45,46]. Specifically, within the $Ln_{0.4}Sr_{0.6}Co_{0.8}Fe_{0.2}O_{3-\delta}$ system (Ln = La, Pr, Nd, Sm, and Gd), Nd-containing compositions exhibited the highest conductivity, followed by La and Pr variants [47]. Consequently, we substituted Sm with Pr (potentially enhancing oxygen permeability) and Y with Sr in Study A's doping scheme, establishing five A-site elements: La, Pr, Nd, Gd, and Sr. Pr, Gd, and Nd were selected not only for their complementary roles in tuning defect chemistry and enhancing transport properties but also for their ability to maintain a stable perovskite structure by optimizing the tolerance factor. This comprehensive approach ensures both high oxygen permeability and excellent operational durability.

This study aims to explore how to maintain the flux of high-flux cobalt-based perovskite films and enhance their stability in a carbon dioxide environment. Achieving high entropy by doping lanthanide elements into single-phase perovskite films may be a viable approach to meet the research objectives [48–52]. In this paper, $La_{0.5}Sr_{0.5}CoO_3$ is used as the parent material. By doping concentrations were systematically adjusted starting from the $La_{0.5}Sr_{0.5}CoO_3$ parent structure. The doping strategy was systematically tailored starting from a $La_{0.5}Sr_{0.5}CoO_3$ parent compound. Precisely controlled amounts of Pr, Nd, and Gd were subsequently incorporated into the A-site with non-equimolar stoichiometries, deliberately increasing the configurational entropy while targeting a non-equimolar, high-entropy cationic distribution, a series of compositions of high-entropy OTMs are synthesized: $La_{0.2}Sr_{0.2}Gd_{0.2}Nd_{0.2}Pr_{0.2}CoO_3$, $La_{0.25}Sr_{0.25}Gd_{0.2}Nd_{0.2}Pr_{0.1}CoO_3$, $La_{0.3}Sr_{0.3}Gd_{0.15}Nd_{0.15}Pr_{0.1}CoO_3$. Structural and oxygen permeation stability are investigated in detail.

## 2. Experimental

### 2.1. Synthesis and characterization

The precursor powders $La_{0.2}Sr_{0.2}Gd_{0.2}Nd_{0.2}Pr_{0.2}CoO_3$ ($LS_{0.2}GNPCoO$), $La_{0.25}Sr_{0.25}Gd_{0.2}Nd_{0.2}Pr_{0.1}CoO_3$ ($LS_{0.25}GNPCoO$), $La_{0.3}Sr_{0.3}Gd_{0.15}Nd_{0.15}Pr_{0.1}CoO_3$ ($LS_{0.3}GNPCoO$), and $La_{0.5}Sr_{0.5}CoO_3$ (LSC) were synthesized using a modified Pechini method. The stoichiometric raw nitrates were weighed and dissolved in distilled water. Subsequently, citric acid was introduced as the chelating agent, while ethylene glycol served as the surfactant.

The molar mass ratio of metal nitrate, citric acid, and ethylene glycol was determined to be 1:2:2. The solution was then placed on a magnetic stirrer and heated and stirred for 8 h until the gel was formed. The resulting gel was dried in an oven at 140 °C for 24 h to obtain a heterogel. The resulting heterogel was ground in a mortar and calcined at 600 °C for 8 h to remove organic impurities. The calcined precursor powder was reground and then recalcined at 950 °C for 10 h to obtain the target powder [26,53]. The phase structure of powders was analyzed by X-ray diffraction (XRD, Rigaku MiniFlex 600) with Cu Kα in the $2\theta$ range of 10-100º at room temperature. XRD data were refined with a Rietveld model, and the crystal structure was made by the VESTA software (Ver. 3.5.7 by Koichi Momma and Fujio Izumi) according to the refinement [54]. The as-prepared powders were ground and then pressed into disks under the pressure of 10 MPa for 5 min, using a circular stainless-steel mold (15 mm in diameter). Dense OTMs were prepared after sintering at 1200 °C for 5 h (with a heating and cooling rate of 1 °C·min$^{-1}$). At last, the OTMs were prepared for testing after being subjected to an ethanol rinse, ethanol effectively removes polishing residue from membrane surfaces while ensuring rapid drying due to its volatility, facilitating subsequent testing procedures. The morphology and element distribution on the surfaces of membranes were illustrated by scanning electron microscopy (SEM, COXEM EM-30AX Plus), backscattered scanning electron microscopy (BSEM), and energy dispersive X-ray spectroscopy (EDS).

**2.2. Oxygen permeation evaluation**

The oxygen permeability of the membranes and their long-term stability under different atmospheres were evaluated using a self-designed experimental setup [55]. All the OTMs were polished into 0.6 mm, which will be sealed at one end of a long corundum tube with high-temperature ceramic glue (Huitian, Hubei, China). The effective working area of the OTMs is approximately 0.95 cm$^2$, calculated according to the inner diameter of the corundum tube and verified through the utilization of a vernier caliper. Synthetic Air (79 vol% $N_2$ and 21 vol% $O_2$) was adopted as the feed gas. Meanwhile, He (99.999 vol% He)/$CO_2$ (99.999 vol% $CO_2$) was used as a sweep gas during the oxygen permeability measurements. All the gas flow rates were tuned by the mass flow meters (MFC, Sevenstar, Beijing, China). The exhaust gases obtained from all reactions were subjected to analysis by gas chromatography (GC, Agilent 7890B, USA) to determine their compositions and concentrations of various gases. The exhaust gas typically comprises permeated oxygen, leaked oxygen, nitrogen, and sweep gas. The permeation rate of oxygen can be calculated using the leakage (eq. (1)) [56]:

$$J_{O_2} = \frac{F}{S} \times \left( C_{O_2} - \frac{C_{N_2}}{4.02} \right) \quad (1)$$

Where $F$ represents the total gas flow entering the GC, while $S$ denotes the effective working area of the OTM. $C_{N_2}$ and $C_{O_2}$ represent the concentrations of $N_2$ and $O_2$ obtained by GC, respectively. The value of 4.02 is based on Knudsen diffusion theory, representing the ratio of $N_2$ to $O_2$ in the leaked gas. It is assumed that the leakage of nitrogen and oxygen through pores or cracks obeys Knudsen diffusion. Given that the dry air used in the experiment consists of a mixture with a nitrogen-to-oxygen volume ratio of 0.79:0.21, the fluxes of leaking nitrogen and oxygen are therefore related by Eq. 2 [57]as follows:

$$J_{N_2}^{Leak} : J_{O_2}^{Leak} = \sqrt{\frac{32}{28}} \times \frac{0.79}{0.21} = 4.02 \quad (2)$$

### 3. Results and discussion

#### 3.1. Phase structure and surface morphology

After calcination in air at 950 °C for 12 h, XRD analysis was conducted to characterize the crystal structure of the $LS_{0.2}GNPCoO$, $LS_{0.25}GNPCoO$, $LS_{0.3}GNPCoO$, and LSC powders. The XRD patterns are shown in Fig. 1. $LS_{0.2}GNPCoO$ powder exhibits an orthorhombic perovskite structure (No. 62: Pbnm), while $LS_{0.25}GNPCoO$, $LS_{0.3}GNPCoO$, and LSC powders show a cubic perovskite structure (No. 221: Pm$\bar{3}$m). The Rietveld refinement model was employed to determine the lattice parameters of these powders, as shown in Table 1.

The selected dopants (Pr, Gd, and Nd) effectively tailor the defect chemistry of the material due to their distinct characteristics and specific substitution behavior at the A-site. Trivalent lanthanide ions (Ln$^{3+}$ = Pr$^{3+}$, Gd$^{3+}$, and Nd$^{3+}$) primarily substitute for Sr$^{2+}$ sites, creating effective positive charge ($Ln_{Sr}^{\bullet}$) according to the defect equation:

$$2Ln^{3+} + 3Sr_{Sr}^{x} \rightarrow 2Ln_{Sr}^{\bullet} + 3Sr^{2+} + V_{Sr}'' \qquad (3)$$

This substitution generates strontium vacancies ($V_{Sr}''$) as the primary charge compensation mechanism, which subsequently influences oxygen vacancy formation through lattice reorganization [58]. The mixed valence nature of Pr enables dynamic redox activity (Pr$^{3+}$ ↔ Pr$^{4+}$ + e$^{-}$), significantly enhancing electronic conductivity via small polaron hopping. This promotes surface oxygen exchange kinetics and facilitates charge transfer processes [26]. Both Gd$^{3+}$ and Nd$^{3+}$, with their stable trivalent states and high Lewis acidity, strengthen metal-oxygen bonds and suppress Sr segregation. Their incorporation improves CO$_2$ tolerance by reducing carbonate formation tendencies while maintaining perovskite stability through optimized tolerance factor. The substitution of La$^{3+}$ by other trivalent lanthanides represents a charge-neutral substitution ($Ln_{Ln}^{x}$), allowing fine-tuning of the ionic radius and oxygen migration pathways without introducing additional charged defects. The synergistic combination of these elements creates a balanced defect chemistry where strontium vacancy formation provides the primary charge compensation, while the multi-cation configuration enhances both structural stability and transport properties.

Eq. (4-5)[43] was employed to compute the mixed entropy ($\Delta S_{mix}$) and the tolerance factor (*t*) of high-entropy perovskites, with the results presented in Table 2. As shown in Table 2, the configurational entropy of the samples increases progressively with higher doping content, among which LS$_{0.2}$GNPCoO exhibits the highest value.

$$\Delta S_{mix} = -R\left[\left(\sum_{a=1}^{n} x_a \ln x_a\right)_{A-site} + \left(\sum_{b=1}^{n} x_b \ln x_b\right)_{B-site} + 3\left(\sum_{c=1}^{n} x_c \ln x_c\right)_{O-site}\right] \quad (4)$$

Where $x_a$, $x_b$, and $x_c$ represent the mole fractions of metallic elements distributed over the A-site, B-site, and incorporated into the anion sublattice (typically as defect species associated with oxygen vacancies or interstitial sites), respectively [43]. The unit of the configurational entropy of mixing ($\Delta S_{mix}$) is J·mol$^{-1}$·K$^{-1}$. This is the standard unit for molar entropy in the International System of Units (SI). This generated controlled entropy enhancement from $\Delta S_{mix}$ = 0.69R (parent) to 1.61R (equimolar pentagonal system), where R is the universal gas constant (8.314 J·mol$^{-1}$·K$^{-1}$). It is the fundamental constant used in thermodynamic equations relating to entropy and energy. The combination of Pr, Gd, and Nd achieves a high mixed entropy ($\Delta S_{mix}$ > 1.5R), which further suppresses cation ordering and phase separation under operational conditions.

$$t = \frac{r_A + r_O}{\sqrt{2}(r_B + r_O)} \quad (5)$$

Where $r_A$, $r_B$, and $r_O$ are the ionic radii of the A-site, B-site, and oxygen ion, respectively. A stable perovskite structure typically requires $t$ to be between 0.8 and 1.0. The ionic radii of the constituent elements directly determine the tolerance factor ($t$), a key geometric descriptor for perovskite stability. In our design, Pr, Gd, and Nd were substituted for a portion of La and Sr at the A-site. This substitution was carefully chosen to maintain the tolerance factor within the stable perovskite range. The introduction of these smaller lanthanides (Pr$^{3+}$, Nd$^{3+}$, and Gd$^{3+}$) moderately decreases the average A-site radius, thereby adjusting the tolerance factor to an optimal value. In addition to the defect chemistry arguments, we have also calculated the tolerance factors ($t$) for the key compositions to demonstrate our rationale for structural stability, and incorporated the $t$ data into Table 2. All designed compositions exhibit $t$ values between 0.95 and 0.97, confirming that the incorporation of Pr, Gd, and Nd at the A-site maintains the perovskite structure in a highly stable state. This optimal geometric condition, combined with the previously discussed defect chemistry, provides a comprehensive foundation for the enhanced oxygen permeation performance observed. This adjustment: Enhances phase stability by reducing internal lattice strain. Promotes oxygen ion transport by creating slightly distorted, yet highly interconnected, oxygen migration pathways. The optimized tolerance factor, combined with the high-entropy effect, results in: Enhanced entropic stabilization of the perovskite phase. Balanced transport properties: The elements Pr, Nd, and Gd were selected for their complementary roles in enhancing electronic and ionic transport properties, respectively, based on established literature reports [47].

The crystal structures of both $LS_{0.25}GNPCoO$ and $LS_{0.3}GNPCoO$ exhibit the same cubic perovskite structure as the LSC parent material. Furthermore, the lattice parameters gradually decrease with increasing doping levels of Nd, Gd, and Pr (Table 1). As is also evident from the Fig. 1, the main peak position (approximately 32°) in the diffraction patterns of the $LS_{0.25}GNPCoO$ and $LS_{0.3}GNPCoO$ sample powders progressively shifts to higher angles with increasing doping content and entropy (Table 2), which corresponds perfectly with the unit cell parameters listed in Table 1. The XRD patterns for each sintered membrane demonstrate that no phase transformation occurred in the sintered membranes compared to the calcined powder. And the surface morphology of the sintered $LS_{0.2}GNPCoO$, $LS_{0.25}GNPCoO$, $LS_{0.3}GNPCoO$, $LS_{0.35}GNPCoO$, and LSC was examined using XRD and SEM (Fig. 2). The SEM images reveal a notably reduced grain size in the high-entropy samples compared to the parent compound. Furthermore, EDXS analysis (Fig. 3) indicates a homogeneous distribution of elements on the surface of the high-entropy samples, whereas Sr segregation is observed in LSC. The elemental ratios detected on all membrane surfaces are close to their theoretical values (Table 3), consistent with the above results.

### 3.2. Oxygen permeation test

The LS$_{0.2}$GNPCoO, LS$_{0.25}$GNPCoO, and LS$_{0.3}$GNPCoO OTMs were tested for oxygen permeability within the temperature ranges of 800 °C to 950 °C under air/He and air/CO$_2$ gradient, respectively. Fig. 4a and b illustrate the relationship between oxygen permeability and temperature. Generally, the oxygen permeation flux $J_{O_2}$ increases with rising temperature, which is attributed to the fact that increasing the temperature can simultaneously enhance both the oxygen surface exchange process and the bulk diffusion process. Fig. 4c compares the $J_{O_2}$ of membranes with different high-entropy ratios at 950 °C. Among these OTMs, the LS$_{0.2}$GNPCoO composition shows the highest $J_{O_2}$ of 1.94 mL·min$^{-1}$·cm$^{-2}$ at 950 °C at air/He gradient. When comparing the oxygen permeation of three high-entropy OTMs at 950°C in a helium atmosphere (Fig. 4a), LS$_{0.2}$GNPCoO, which possesses the largest lattice parameters, exhibits the highest flux.

We also explored the differences in performance and stability under pure He or pure CO$_2$ purging. Since CO$_2$ has a stronger adsorption effect on the surface of metal oxides, the oxygen exchange sites on the membrane surface are reduced[59]. Therefore, after switching the purge gas to CO$_2$, $J_{O_2}$ decreases. The activation energy ($E_a$) calculated based on the Arrhenius plot is consistent with the aforementioned results (Fig. 6). When purged with CO$_2$, all membranes require a higher $E_a$ to separate oxygen. Compared with the other two membranes, LS$_{0.25}$GNPCoO has the least increase in $E_a$ and the smallest decrease in oxygen permeability.

The higher activation energy ($E_a$) for LS$_{0.3}$GNPCoO under CO$_2$ is attributed to a more negative entropy of activation ($\Delta S^{\ddagger}$) for the surface oxygen exchange reaction. The LS$_{0.2}$GNPCoO composition possesses a higher A-site configurational entropy. This creates a more disordered and flexible surface that better stabilizes the constrained transition state required for carbonate (CO$_3^{2-}$) formation during the surface reaction with CO$_2$. This results in a less negative $\Delta S^{\ddagger}$. In contrast, the surface of the La/Sr-rich composition is more structurally rigid due to its lower configurational entropy. It is less able to accommodate the ordered transition state, leading to a more negative $\Delta S^{\ddagger}$. According to Eq. 6[60], a more negative $\Delta S^{\ddagger}$ directly contributes to a higher apparent $E_a$.

$$E_a = \Delta H^{\ddagger} - T\Delta S^{\ddagger} \qquad (6)$$

When the purge gas is changed from He to $CO_2$ at 950 °C, the optimal OTM shifts from $LS_{0.2}$GNPCoO to $LS_{0.25}$GNPCoO, and the oxygen permeability of $LS_{0.25}$GNPCoO in a $CO_2$ atmosphere can reach 1.46 mL·min$^{-1}$·cm$^{-2}$. Moreover, high-entropy modification can also prevent the formation of impurities such as carbonates and enhance the structural stability of the membrane in a high-temperature $CO_2$ atmosphere. In the long-term oxygen permeability test at 950 °C (Fig. 4c), after 40 h of He purging and 100 h of $CO_2$ purging, the $J_{O_2}$ of the high-entropy membranes showed no downward trend, demonstrating excellent performance stability in high-temperature and harsh environments.

The difference in oxygen permeation behavior between the orthogonal $LS_{0.2}$GNPCoO and the cubic $LS_{0.25}$GNPCoO membranes under different sweep gases can be explained through the combined analysis of their crystal phase, tolerance factor, and lattice parameters, which collectively determine the dominant oxygen transport mechanism in each environment. Our XRD refinements confirm that $LS_{0.2}$GNPCoO crystallizes in an orthorhombic perovskite structure, whereas $LS_{0.25}$GNPCoO maintains a cubic structure. Despite this difference, the calculated tolerance factors ($t$) for both compositions fall within the stable perovskite range (0.95–0.97). The orthorhombic distortion in $LS_{0.2}$GNPCoO indicates a slightly higher degree of lattice distortion, which influences its transport properties.

Lattice Parameters and Bulk Diffusion (Dominant under He Sweep): As indicated by the refinement results in Table 1 and Fig. 1, the lattice parameters of $LS_{0.2}$GNPCoO are larger than those of both $LS_{0.25}$GNPCoO and $LS_{0.3}$GNPCoO. $LS_{0.2}$GNPCoO (Orthorhombic), despite its lower symmetry, exhibits larger effective lattice parameters and unit cell volume compared to the cubic compositions. This expanded lattice can create more spacious, albeit anisotropic, channels for oxygen ion migration. Under inert He sweep, where surface exchange is facile, oxygen permeation is primarily governed by bulk diffusion. The larger lattice dimensions of the orthorhombic phase appear to facilitate this process more effectively, resulting in the highest flux for $LS_{0.2}GNPCoO$ in this environment [45].

Surface Chemistry and Phase Stability (Dominant under $CO_2$ Sweep): Under $CO_2$ sweep, the permeation process becomes limited by surface exchange kinetics. $CO_2$ poisons the membrane surface by forming carbonates (e.g., $SrCO_3$), which block active sites. The cubic structure of $LS_{0.25}GNPCoO$ possesses higher symmetry and isotropic ionic transport pathways. More importantly, its specific non-equimolar A-site cation ratio (La, Sr, Gd, Nd, and Pr) appears to optimize the surface chemistry. We propose that this composition achieves a superior balance of Lewis acid-base characteristics, which suppresses Sr segregation and consequently inhibits carbonate formation. Furthermore, its cubic phase and high configurational entropy enhance the phase stability against surface decomposition under $CO_2$ attack, ensuring the surface remains catalytically active for oxygen exchange [61,62]. For example, oxygen ions can easily migrate on the *ab* plane while experiencing difficulty along the *c*-axis in the orthorhombic $La_{0.64}(Ti_{0.92}Nb_{0.08})O_{2.99}$ compound [63].

Therefore, the performance divergence arises from a shift in the rate-limiting step influenced by crystal structure: He sweep: Bulk diffusion is limiting, Orthorhombic $LS_{0.2}GNPCoO$ excels due to its larger lattice dimensions, facilitating ion migration. $CO_2$ sweep: Surface exchange is limiting, Cubic $LS_{0.25}GNPCoO$ excels due to its superior surface stability and entropy-enhanced resistance to $CO_2$ poisoning.

### 3.3. Stability of materials

To initially investigate the structural stability of high-entropy materials, we subjected the powder samples to heat treatment in pure $CO_2$ for 24 h before sintering. The XRD results indicate that no phase transformation occurred in the high-entropy oxides treated in $CO_2$ at 800 °C, 850 °C, and 900 °C (Fig. 5). Moreover, it can be observed that as the doping precisely controlled amounts of Nd/Gd/Pr into the A-site with non-equimolar stoichiometries of $La_{0.5}Sr_{0.5}CoO_3$. This generated controlled entropy enhancement from $\Delta S_{mix}$ = 0.69R (parent) to 1.61R (equimolar quinary system). As revealed in Fig. 5, carbonate suppression is positively correlated with entropy, demonstrating that configurational entropy dominates degradation resistance under $CO_2$ exposure. Notably, $LS_{0.2}$GNPCoO and $LS_{0.25}$GNPCoO are almost identical to the fresh powder, which is in stark contrast to the non-high-entropy parent material. Based on the combined analysis of Fig. 5 and Table 4, the formation of $SrCO_3$ shows a markedly non-linear relationship with Sr content, highlighting the distinct advantage of high-entropy design. While the Sr content decreases by 50 % (from 50 % in LSC to 25 % in $LS_{0.25}$GNPCoO), the corresponding $SrCO_3$ formation at 900°C decreases disproportionately by nearly 92 % (from 21.5 % to 1.75 %). This dramatic reduction, far exceeding what would be expected from merely halving the Sr content, provides compelling evidence that the high-entropy configuration itself significantly suppresses carbonate formation, rather than the effect being solely due to the reduction in Sr content. Thus, increasing the mixed entropy of the membrane can reduce the formation of carbonates, thereby preventing the reduction in oxygen permeability and structural damage of Sr-containing OTMs [64,65]. And as shown in Fig. 5, the orthorhombic $LS_{0.2}$GNPCoO demonstrated stability against carbonate formation that was at least equivalent to its cubic counterpart, $LS_{0.25}$GNPCoO, in a $CO_2$ atmosphere. This reveals that effective suppression of Sr segregation is not exclusive to the cubic phase but can also be achieved in an orthorhombic structure through tailored composition.

Additionally, the performance stability of the long-term oxygen permeability test is shown in Fig. 4c. The three high-entropy OTMs stably operated for 40 h and 100 h under He and $CO_2$ sweeping at 950 °C, with virtually no performance decline. The high-entropy OTMs exhibited only a slight performance drop after switching. This may be attributed to the fact that carbon dioxide partially occupies the surface active sites, rather than carbonates disrupting the oxygen transport pathways. Figs. 7 and 8 display the SEM-EDXS images of the sweep side of the spent membranes after the long-term oxygen permeability test. The morphological changes on the $CO_2$ sweep side of the spent membranes represent a central aspect of our degradation analysis, and two concurrent processes have been identified: Surface Carbonation, the low oxygen partial pressure ($pO_2$) and high $CO_2$ concentration on the sweep side drive the segregation of Sr cations and their subsequent reaction with $CO_2$ to form $SrCO_3$. This is evident in Fig. 8, where the Sr elemental mapping shows the formation of minor granular or plate-like agglomerates [27]. And thermal Microstructural Evolution, prolonged annealing at high temperature also leads to grain coarsening and surface densification. No second-phase impurity grains were found on the surface of all high-entropy membranes, and there was no significant segregation in some aspects of the element maps. This indicates that three OTMs maintained the stability of the phase structure during the test, which is conducive to the oxygen permeation process. The element content of the sweep side, characterized by EDXS (Table 5), is essentially consistent with that of the fresh membrane, indicating no large-scale cation transfer between the material surface and interior, and demonstrating the excellent performance stability of the high-entropy OTM in a high-temperature $CO_2$-containing working atmosphere. This provides direct visual evidence that the high-entropy strategy effectively suppresses cation segregation, thereby enhancing phase stability and extending membrane lifetime under operating conditions.

## 4. Conclusions

In this study, we designed and synthesized a series of novel high-entropy cobalt-based perovskite oxygen transport membranes (OTMs), designated as $LS_{0.2}GNPCoO_3$, $LS_{0.25}GNPCoO_3$, and $LS_{0.3}GNPCoO_3$, through multi-lanthanide doping at the A-site. The introduction of configurational entropy significantly improved the materials' tolerance to $CO_2$ and reducing atmospheres, while retaining high oxygen permeability.

Among the compositions, $LS_{0.25}GNPCoO_3$ demonstrated optimal performance with oxygen permeation fluxes reaching 1.62 mL·min$^{-1}$·cm$^{-2}$ under He and 1.46 mL·min$^{-1}$·cm$^{-2}$ under $CO_2$ sweep at 950 °C. It also exhibited exceptional operational stability, showing no performance degradation over 100 hours in a $CO_2$ environment and minimal carbonate formation compared to the parent material $La_{0.5}Sr_{0.5}CoO_3$. The phase demonstrated exceptional stability after exposure to the $CO_2$ atmosphere at 800-900 °C.

This work underscores that while high configurational entropy enhances structural and chemical stability, the highest functional performance also depends on a balanced optimization of ionic transport, surface kinetics, and crystal structure. The $LS_{0.25}GNPCoO_3$ composition exemplifies this balance, showing superior properties without possessing the maximum entropy value. These findings highlight the potential of entropy-stabilized OTMs for use in membrane reactors and high-temperature electrochemical devices under challenging operational conditions.

**Acknowledgments**

This work is supported by the National Natural Science Foundation of China (12404165, 12274471, 11922415), Guangdong Basic and Applied Basic Research Foundation (Grant No.

2025A1515010311), Guangzhou Science and Technology Programme (No. 2024A04J6415), the State Key Laboratory of Optoelectronic Materials and Technologies (Sun Yat-Sen University, No. OEMT-2024-ZRC-02), Key Laboratory of Magnetoelectric Physics and Devices of Guangdong Province (Grant No. 2022B1212010008), and Research Center for Magnetoelectric Physics of Guangdong Province (2024B0303390001). Lingyong Zeng acknowledges the Postdoctoral Fellowship Program of CPSF (GZC20233299) and the Fundamental Research Funds for the Central Universities, Sun Yat-sen University (24qupy092).

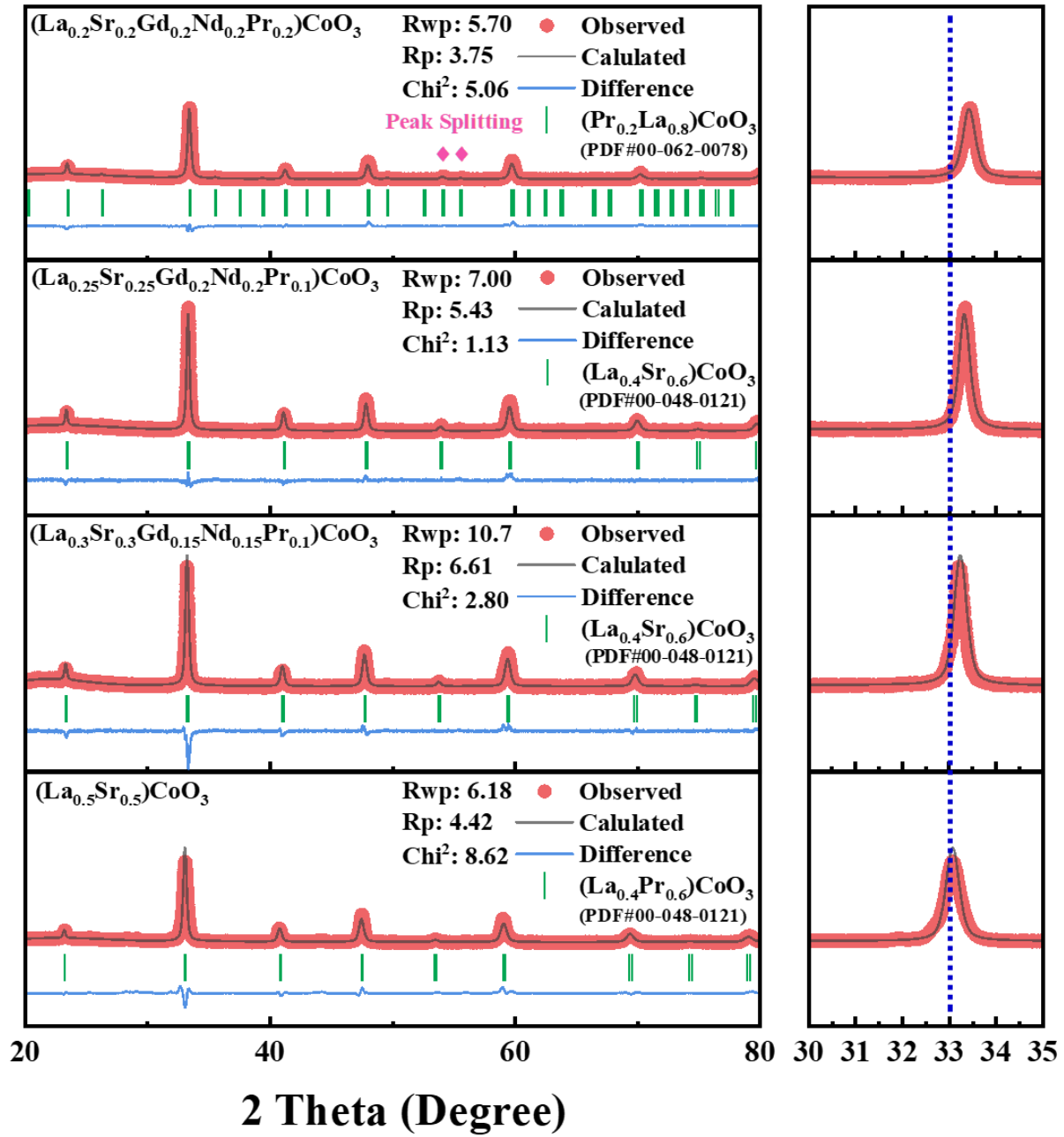

Fig. 1. XRD refinements of LS$_{0.2}$GNPCoO, LS$_{0.25}$GNPCoO, LS$_{0.3}$GNPCoO and LSC powders after 950 °C for 10 h calcination.

Table 1. Cell parameters of LS$_{0.2}$GNPCoO, LS$_{0.25}$GNPCoO, LS$_{0.3}$GNPCoO and LSC powders calcined at 950 °C obtained by XRD data refinement with the Rietveld model.

|  | Perovskite | | | | | | |
| --- | --- | --- | --- | --- | --- | --- | --- |
|  | a/Å | b/Å | c/Å | Rwp | Rp | Chi$^2$ | |
| LS$_{0.2}$GNPCoO | 5.35525 | 5.36979 | 7.57736 | 5.70 | 3.75 | 5.06 | (No. 62: Pbnm) |
| LS$_{0.25}$GNPCoO | 3.80185 | 3.80185 | 3.80185 | 7.00 | 5.43 | 1.13 | |
| $LS_{0.3}GNPCoO$ | 3.81073 | 3.81073 | 3.81073 | 10.7 | 6.61 | 2.80 | (*No.* 221: Pm$\bar{3}$m) |
| *LSC* | 3.83016 | 3.83016 | 3.83016 | 6.18 | 4.42 | 8.62 | |

Table 2. Mixed entropy($\Delta S_{mix}$) and tolerance factors (*t*) of LS0.2GNPCoO, LS0.25GNPCoO, LS0.3GNPCoO and LSC.

| Materials | $\Delta S_{mix}$ | *t* |
|---|---|---|
| $LS_{0.2}GNPCoO$ | 1.609438 | 0.958 |
| $LS_{0.25}GNPCoO$ | 1.567181 | 0.961 |
| $LS_{0.3}GNPCoO$ | 1.425651 | 0.966 |
| *LSC* | 0.693147 | 0.985 |

Table 3. Experimental and theoretical values of the atomic ratio of each element on the surface of fresh single-phase membranes.

| Composition | State | Element (Atomic percentage) | | | | | |
|---|---|---|---|---|---|---|---|
| | | La | Sr | Gd | Nd | Pr | Co |
| $LS_{0.2}GNPCoO$ | *Theoretical* | 10.00% | 10.00% | 10.00% | 10.00% | 10.00% | 50.00% |
| | *Experimental* | 10.34% | 8.47% | 9.38% | 10.02% | 9.68% | 52.10% |
| $LS_{0.25}GNPCoO$ | *Theoretical* | 12.5% | 12.5% | 10.00% | 10.00% | 5.00% | 50.00% |
| | *Experimental* | 12.24% | 11.89% | 9.39% | 9.77% | 4.58% | 52.13% |
| $LS_{0.3}GNPCoO$ | *Theoretical* | 15.00% | 15.00% | 7.50% | 7.50% | 5.00% | 50.00% |
| | *Experimental* | 14.77% | 14.55% | 6.89% | 7.01% | 4.41% | 52.37% |
| *LSC* | *Theoretical* | 25.00% | 25.00% | | | | 50.00% |
| | *Experimental* | 23.25% | 24.72% | | | | 52.04% |

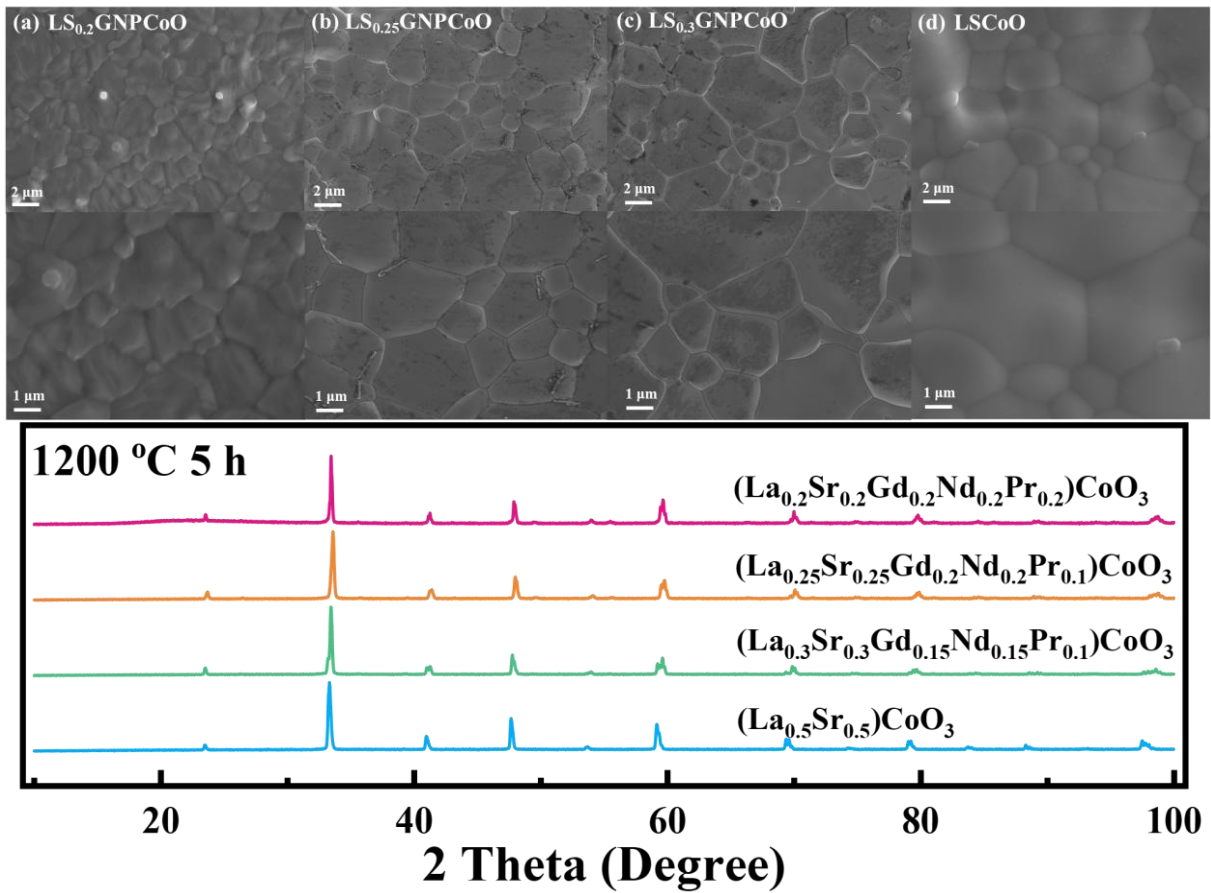

Fig. 2. (Upper panel) SEM images showing the surface morphology of fresh (a) LS$_{0.2}$GNPCO, (b) LS$_{0.25}$GNPCO, (c) LS$_{0.3}$GNPCO, and (d) LSC membranes sintered at 1200 °C for 5 hours. (Lower panel) The corresponding XRD patterns of the membrane surfaces.

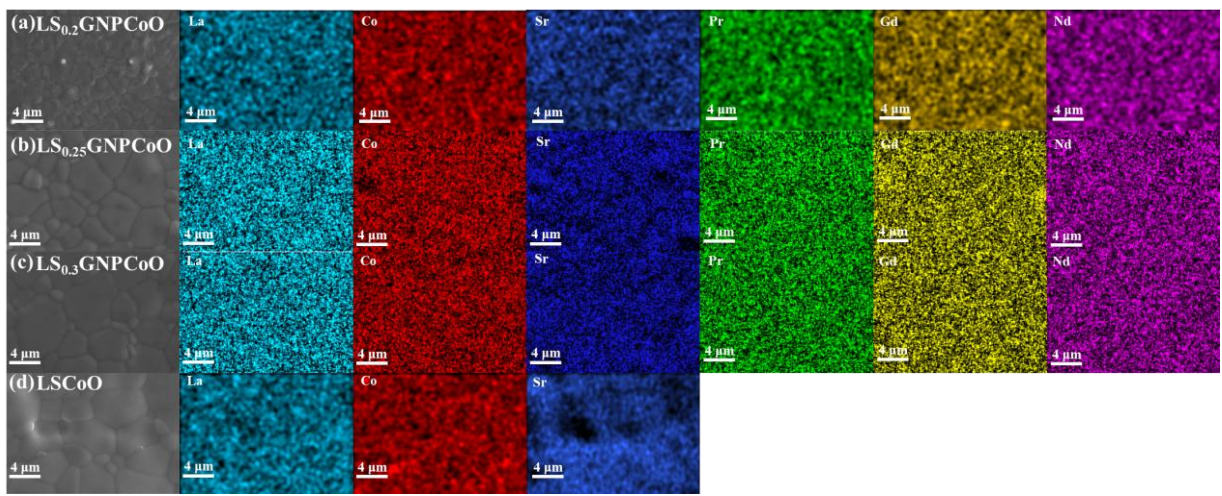

Fig. 3. EDXS images of elements on the surface of LS$_{0.2}$GNPCoO, LS$_{0.25}$GNPCoO, LS$_{0.3}$GNPCoO and LSC fresh membranes sintered at 1200 °C for 5 h.

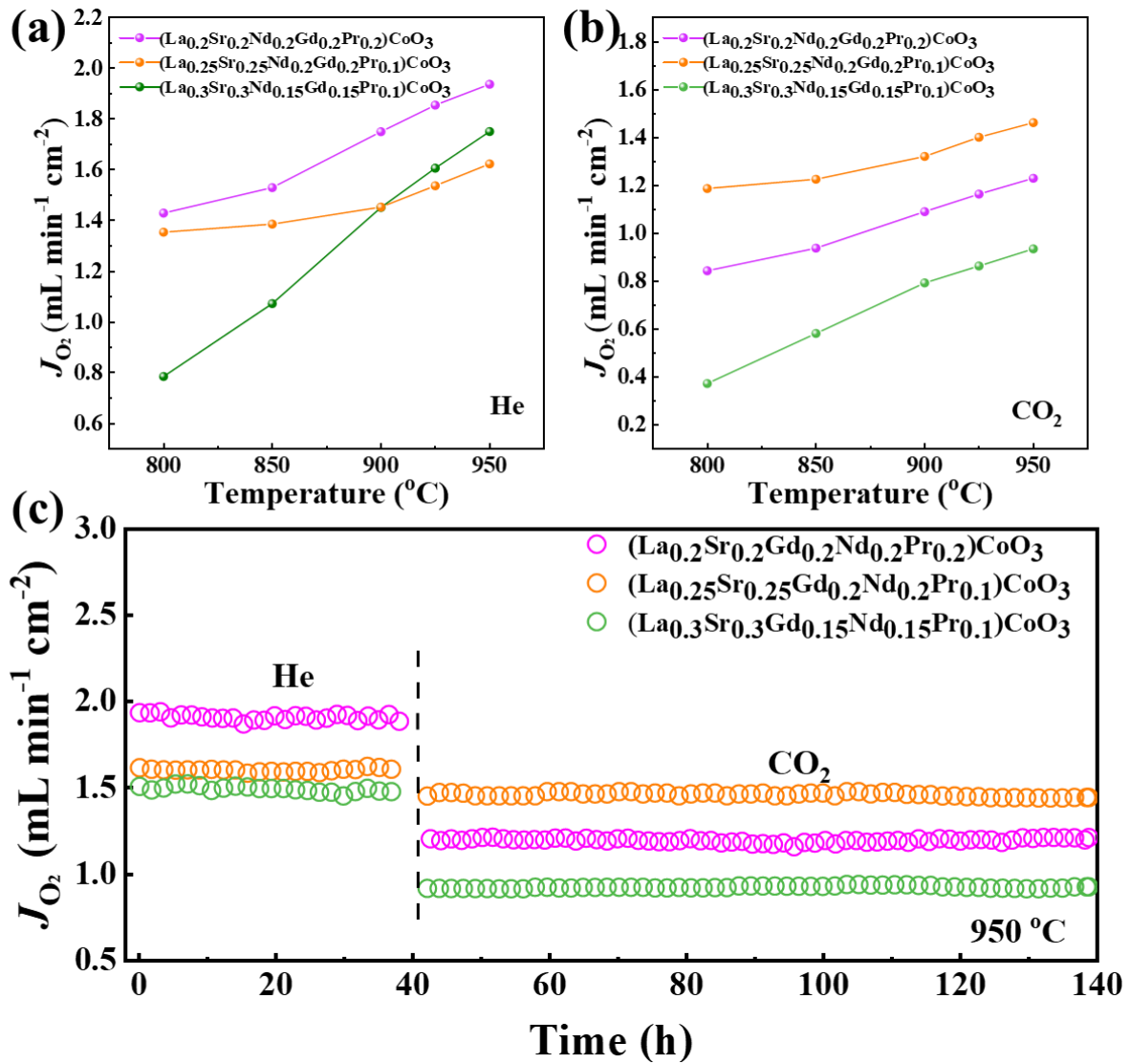

Fig. 4. Curves of oxygen permeation fluxes of LS$_{0.2}$GNPCoO, LS$_{0.25}$GNPCoO, LS$_{0.3}$GNPCoO membranes versus temperature with (a) He and (b) CO$_2$ sweeping. (c) Oxygen permeation fluxes during long-term tests at 950 °C.

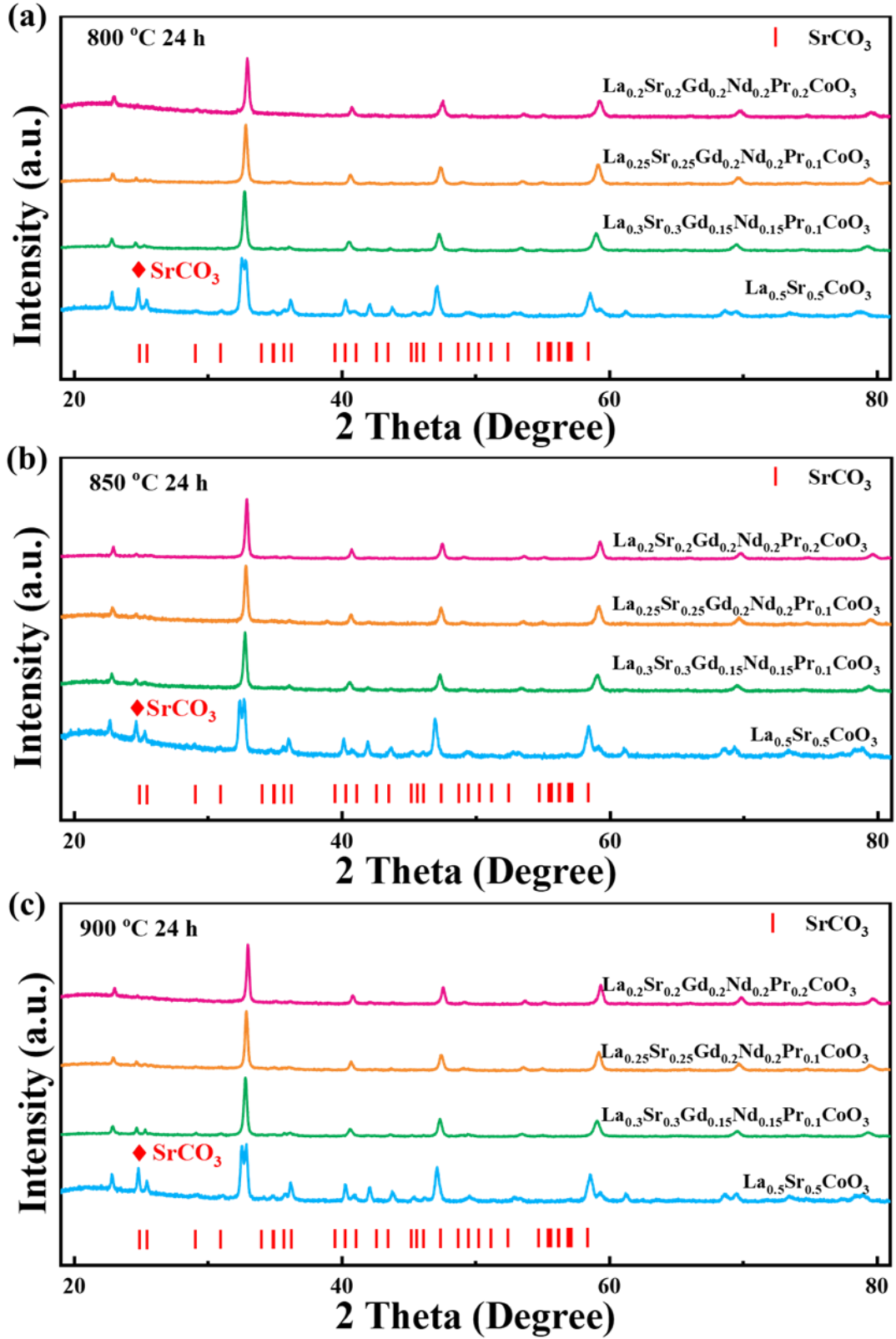

Fig. 5. XRD patterns of LS$_{0.2}$GNPCoO, LS$_{0.25}$GNPCoO, LS$_{0.3}$GNPCoO, and LSC powders treated in pure CO$_2$ at (a) 800 ℃, (b) 850 ℃, and (c) 900 ℃ for 24 hours.

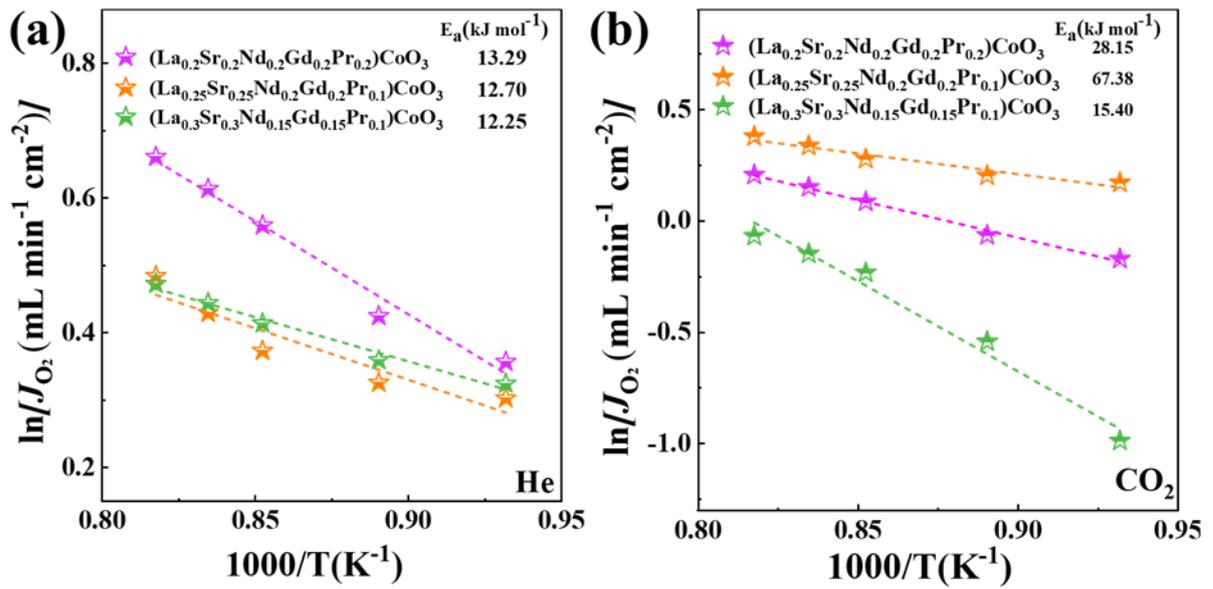

Fig. 6. Arrhenius plots of $J_{O_2}$ and activation energy fitting of LS$_{0.2}$GNPCoO, LS$_{0.25}$GNPCoO, LS$_{0.3}$GNPCoO OTMs with (a) He and (b) CO$_2$ sweeping.

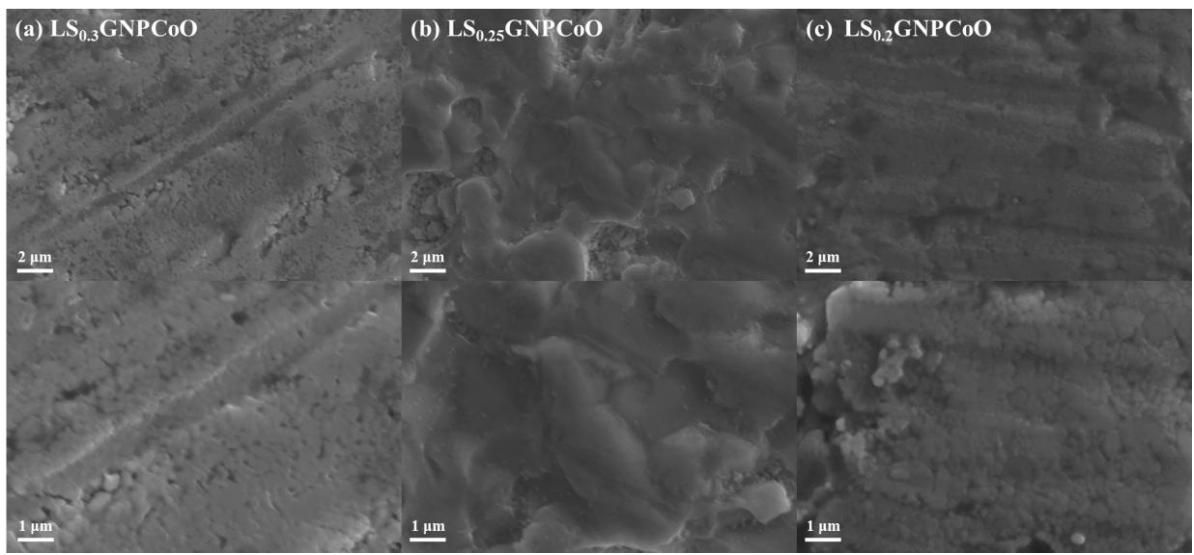

Fig.7. Surface topography of the sweep side of spent LS$_{0.2}$GNPCoO, LS$_{0.25}$GNPCoO, and LS$_{0.3}$GNPCoO OTMs after the test.

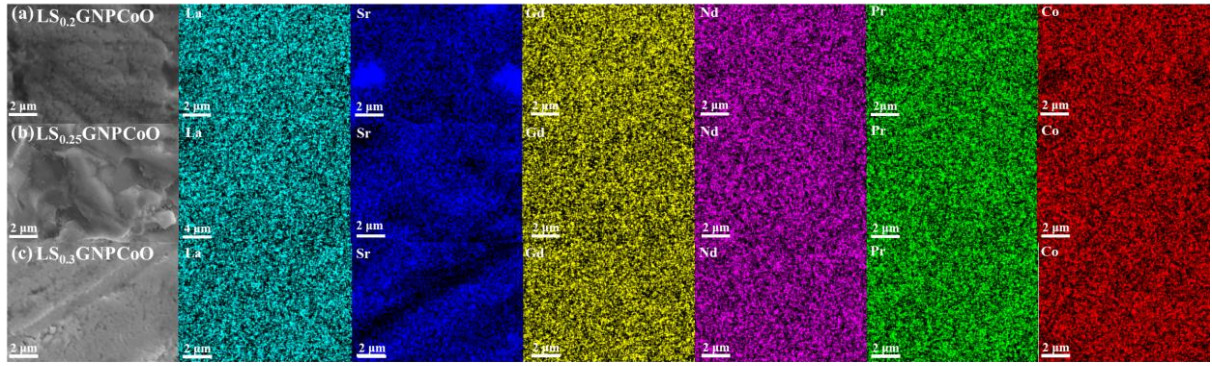

Fig. 8. Element distribution on the sweep side of spent (a)LS$_{0.2}$GNPCoO, (b)LS$_{0.25}$GNPCoO, and (c)LS$_{0.3}$GNPCoO OTMs after the test.

Table 4. Relationship between Sr content and SrCO$_3$ formation in various powder samples after 24 h exposure to CO$_2$ atmosphere at different temperatures.

| Composition | Sr Content (mol%) | SrCO$_3$ Content at 800 °C | SrCO$_3$ Content at 850 °C | SrCO$_3$ Content at 900°C |
|---|---|---|---|---|
| *LSC* | 50 % | 19.6 % | 18.1 % | 21.5 % |
| *LS$_{0.3}$GNPCoO* | 30 % | 1.77 % | 2.20 % | 3.78 % |
| *LS$_{0.25}$GNPCoO* | 25 % | <1 % | <1 % | 1.75 % |
| *LS$_{0.2}$GNPCoO* | 20 % | <1 % | <1 % | <1 % |

Table 5. Surface Atomic Percentages (at%) of Elements in Fresh vs. Spent Membranes.

| Composition | State | Element (Atomic percentage) | | | | | |
|---|---|---|---|---|---|---|---|
| | | La | Sr | Gd | Nd | Pr | Co |
| *LS$_{0.2}$GNPCoO* | *Fresh* | 10.34% | 8.47% | 9.38% | 10.02% | 9.68% | 52.10% |
| | *Spent* | 9.85% | 10.02% | 9.40% | 9.66% | 9.36% | 51.71% |
| *LS$_{0.25}$GNPCoO* | *Fresh* | 12.24% | 11.89% | 9.39% | 9.77% | 4.58% | 52.13% |
| | *Spent* | 11.97% | 11.86% | 9.34% | 9.61% | 4.69% | 52.53% |
| *LS$_{0.3}$GNPCoO* | *Fresh* | 14.77% | 14.55% | 6.89% | 7.01% | 4.41% | 52.37% |
| | *Spent* | 14.79% | 14.15% | 6.98% | 7.34% | 7.12% | 52.11% |